\documentclass[aps,jcp,twocolumn]{revtex4}
\usepackage{graphicx}% Include figure files
\usepackage{dcolumn}% Align table columns on decimal point
\usepackage{bm}% bold math
\usepackage{amsmath}% bold math
\makeatletter
\def\@dotsep{4.5}
\makeatother

\begin{document}
\title{Kinetic Monte Carlo Studies of Hydrogen Abstraction from Graphite }
\author{H.M. Cuppen}
\affiliation{Leiden Observatory, Leiden University, P.O. Box 9513, 2300 RA Leiden, The Netherlands}
\author{L. Hornek{\ae}r}
\affiliation{Dept. Physics and Astronomy and Interdisciplinary Nanoscience Centre (iNANO), University of Aarhus, DK-8000 Aarhus C, Denmark}
%-------------------------------------------------------------------------
\date{\today}
\begin{abstract}
We present Monte Carlo simulations on Eley-Rideal abstraction reactions of atomic hydrogen chemisorbed on
graphite. The results are obtained via a hybrid approach using energy barriers derived from DFT calculations as
input to Monte Carlo simulations. By comparing with experimental data we discriminate between contributions from
different Eley-Rideal mechanisms. A combination of two different mechanisms yields, good quantitative and
qualitative agreement between the experimentally derived and the simulated Eley-Rideal abstraction cross sections
and surface configurations. These two mechanisms include a direct Eley-Rideal reaction with fast diffusing H
atoms and a dimer mediated Eley-Rideal mechanism with increased cross section at low coverage. Such a dimer
mediated Eley-Rideal mechanism has not previously been proposed and serves as an alternative explanation to the
steering behavior often given as the cause of the coverage dependence observed in Eley-Rideal reaction cross
sections.
\end{abstract}

\maketitle
\section{Introduction}
Molecular hydrogen is the most abundant molecule in the interstellar medium (ISM), where it serves as an
important coolant and as a precursor for the formation of more complex molecules. In the cold and dilute
interstellar medium no efficient gas phase routes exist for the formation of H$_2$. Hence, the most likely route
is to form the molecule on the surfaces of small dust particles. This process has been intensely studied over the
past years \cite{Pirronello:1997,Pirronello:1997a, Pirronello:1999, Hornekaer:2003, Hornekaer:2005, Perry:2003,
Cazaux:2004, Chang:2005, Cuppen:heating, Perets:2005, Katz:1999, Dulieu:2005}. The general consensus is that
molecular hydrogen can be formed by a diffusive Langmuir-Hinshelwood mechanism, in the very cold regions below
$\sim$20 K \cite{Hollenbach:1971}. At these low temperatures the hydrogen atoms are physisorbed on the dust
particles, but can still easily move on the surface. Once two atoms meet, they will react to form molecular
hydrogen. If the surface temperature becomes too high, the residence time of atoms on the surface becomes so
short that, at the low fluxes in the ISM, the chance of two atoms meeting becomes negligible. Observations show
that molecular hydrogen is also formed in warmer areas ($>$ 20 K) like Photon Dominated Regions (PDRs) and
post-shock regions. A possible mechanism at these conditions would be through an Eley-Rideal reaction where an
incoming hydrogen atom reacts with another H atom that is chemically bound to the surface. Since a considerable
fraction of the interstellar grains is expected to consist of carbonaceous material, a popular model system for
this process is hydrogen abstraction from graphite. This Eley-Rideal reaction has been studied by Density
Functional Theory (DFT) and Quantum Wave packet calculations \cite{Farebrother:2000, Meijer:2001, Sha:2002,
Sha:2002a, Morisset:2004, Martinazzo:2006}. Based on these calculations several different mechanisms have been
proposed to contribute to the Eley-Rideal abstraction process. These include: Direct Eley-Rideal
\cite{Farebrother:2000, Meijer:2001, Sha:2002, Sha:2002a, Morisset:2004, Martinazzo:2006}, barrier-less
abstraction of one hydrogen atom forming part of a para-dimer configuration \cite{Bachellerie:2007} and
abstraction by rapidly diffusing H atoms in physisorbed states \cite{Bonfanti:2007}.

However, due to computational limitations such detailed calculations are unable to take the full complexity of
the H-graphite system into account. Hence, in the calculations the flat surface approximation is often used and
generally either none or only a single C atom on the simulated graphite surface were allowed to relax during the
Eley-Rideal reaction. These simplifications are potentially problematic. By not allowing for relaxation of the C
atoms on the surface zero barrier reaction channels, such as sticking of hydrogen atoms into specific dimer
configuration \cite{Hornekaer:2006II,Rougeau:2006}, are in some cases artificially closed, making it impossible
to evaluate the contributions from different proposed mechanisms to the Eley-Rideal abstraction process. Eg.
calculations on the Eley-Rideal reaction by Martinazzo and Tantardini \cite{Martinazzo:2006} show a higher
probability for reaction if the incoming H atom is 0.5 - 1.5 \AA~away from the target H atom instead of a direct
hit on. This distance corresponds to the H-H distance in an ortho-dimer configuration (see Fig. 1) on the
graphite surface (1.42 \AA). Hence, surface corrugation and competing sticking reactions must be expected to
influence the abstraction behavior. In these calculations, however, dimer formation was not a possible route
since the flat surface approximation was used and only the carbon atom underneath the original hydrogen atom was
allowed to relax. Impact parameters corresponding to the para-dimer (see Fig. 1) distance (2.84 \AA) were not
included in the study.

\begin{figure}
\begin{center}
\includegraphics{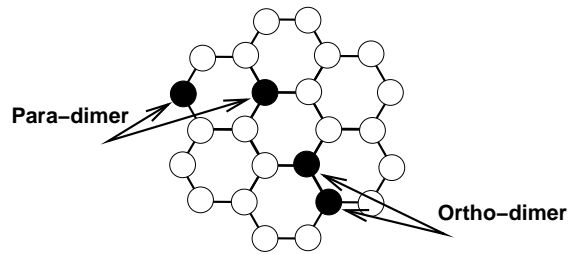}
\end{center}
\caption{The two stable dimer configurations found by Hornek{\ae}r et al. \cite{Hornekaer:2006I}.}
\label{dimers}
\end{figure}

Both experimental observations and theoretical calculations show that hydrogen atoms preferentially form
adsorbate clusters on the graphite surface. Scanning Tunneling Microscopy (STM) investigations show that at a
hydrogen atom coverage of $\sim$ 0.5 $ \%$ more than 75 $\%$ of all surface configurations are clusters
\cite{Hornekaer:2006II}. This indicates that $\sim$ 85 $\%$ of the hydrogen atoms are part of larger clusters. In
particular, the existence of hydrogen dimer configurations has been studied theoretically
\cite{Ferro:2003,Miura:2003,Hornekaer:2006I,Hornekaer:2006II,Rougeau:2006} and experimentally
\cite{Hornekaer:2006I, Andree2006}. Hornek{\ae}r et al. \cite{Hornekaer:2006I} identified two stable hydrogen
dimer configurations on the graphite surface, an ortho-dimer and a para-dimer (see Fig.~\ref{dimers}). DFT
calculations show that formation of the para-dimer is barrierless and that formation of the ortho-dimer has a
reduced barrier \cite{Hornekaer:2006II}, which makes dimer formation a competing channel to Eley-Rideal
abstraction at non-zero impact parameter.

One set of experiments on the Eley-Rideal abstraction reaction forming HD has been reported \cite{Zecho:2002,
Zecho:2002a}. In these experiments a high cross section for the abstraction reaction was observed which varied
from 17 \AA$^2$ at low coverage to 4 \AA $^2$ at high coverage. This variation in the cross section with coverage
has been ascribed to a steering effect \cite{Sha:2002, Sha:2002a}. However, other mechanisms could also
contribute to the high cross section at low coverage.  Options include the barrier-less abstraction of one of the
H atoms in a para-dimer configuration \cite{Bachellerie:2007} and abstraction by fast diffusing H atoms
\cite{Bonfanti:2007}. We propose a further possibility, namely abstraction via a hydrogen dimer state, where the
incoming atom is not immediately thermalized and some of its excess initial energy is used to overcome the
barrier to H$_2$ formation and desorption. The finding that adsorption into the hydrogen para-dimer state is
barrierless \cite{Hornekaer:2006II} make this a strong competing reaction channel.

Hence, in total 5 different abstraction mechanisms have been proposed to contribute to Eley-Rideal abstraction of
hydrogen on graphite:

\begin{enumerate}
\item direct Eley-Rideal \cite{Farebrother:2000,Morisset:2004}.
\item direct Eley-Rideal with steering \cite{Zecho:2002,Sha:2002,Sha:2002a}.
\item preferred direct Eley-Rideal with hydrogen atoms part of a para-dimer \cite{Bachellerie:2007}.
\item a dimer mediated reaction where the incoming atom is first adsorbed into a dimer configuration and, before
thermalizing to the substrate temperature, reacts with the other atom in the dimer to form H$_2$ and desorb.
\item Eley-Rideal reactions by fast diffusing H atoms in the physisorption state \cite{Bonfanti:2007}.
\end{enumerate}

However, due to the simplifications needed in the complex DFT and quantum wave packet calculations a quantitative
comparison between experimental data and theory has not been possible. In this paper we employ a hybrid approach
in which we include the findings of ab-initio DFT calculations in a Monte Carlo simulation program and then
simulate Eley-Rideal abstraction experiments on a more realistic surface area with more complex hydrogen
adsorbate configurations than what is possible in the DFT and wave packet calculations. Through this quantitative
approach we aim to discriminate between the contributions of the different underlying mechanisms for hydrogen
abstraction.

\section{Monte Carlo model}
\begin{figure}
\begin{center}
\includegraphics{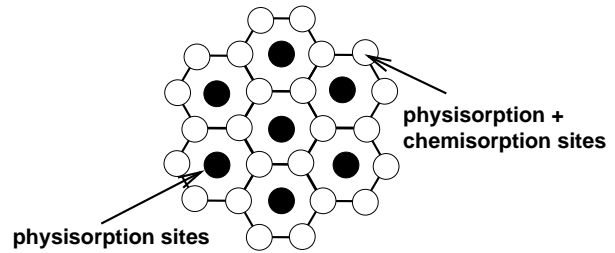}
\end{center}
\caption{The adsorption sites used in the Monte Carlo simulations.}
\label{system}
\end{figure}

The Monte Carlo simulation program is a so-called lattice-gas simulation program, where the atoms are confined to
an adaptive grid depending on the initial coverage. The hydrogen atoms can chemisorb to the graphite surface at
the sites directly on top of the carbon atoms as indicated in Figure \ref{system}. The puckering of the carbon
atom upon chemisorption is included indirectly by using the barriers for chemisorption and binding energies from
DFT calculations that allow this motion. Interaction between two hydrogen atoms is accounted for in a similar way
as will be discussed in the following sections. The hydrogen atoms can physisorb to the surface without a
barrier. Since the potential energy surface for physisorption is rather flat, there are probably no specific
physisorption site, however since the lattice-gas model forces us to choose specific sites, we confine
physisorption to the sites directly above the carbon atoms and an additional site at the center of the ring
(Figure \ref{system}).

The simulation starts with a clean graphite surface. The first event will be a deposition attempt of the first
atom. The time at which this attempt will occur is
\begin{equation}
t_{\rm dep} = -\frac{\ln\left(X\right)\sigma}{f} + t
\label{tdep}
\end{equation}
where $\sigma$ is the density of sites, $X$ is a random number between 0 and 1, $f$ is the hydrogen flux in atoms
per time per area, and $t$ is the current time. Deposition times for subsequent sticking events are determined by
the same expression (Eq.~\ref{tdep}). How the atoms will bind to the surface is determined by another random
number and depends on the barrier for chemisorption in that specific position. If the site is already occupied by
another hydrogen atom, an abstraction reaction is considered by comparing a Boltzmann factor including the
abstraction barrier against a random number. Upon reaction both atoms will leave the surface in the form of
H$_2$, else the incoming atom is deflected. The abstraction channel can be closed by making the barrier
infinitely large.

Once hydrogen atoms populate the surface, they can diffuse, desorb, or recombine with other atoms to form H$_2$.
For each surface hydrogen the time at which they will undergo one of these events is determined by
\begin{equation}
t^{i} = -\frac{\ln\left(X\right)}{R_{\rm dif}^i + R_{\rm des}^i + R_{\rm rec}^i} + t\label{t^i}
\end{equation}
with $R_{\rm dif}^i$, $R_{\rm des}^i$, and $R_{\rm rec}^i$ the rate for diffusion, desorption, and reaction of
atom $i$, respectively. Another random number determines which of the three events occur according to their
relative probability of occurrence. Reaction between two chemisorbed atoms can only occur if they form a dimer
configuration. The rates are given by
\begin{equation}
R = \nu \exp\left(-\frac{E}{k T'}\right),
\label{R}
\end{equation}
where $\nu$ is the attempt frequency which is assumed to be 10$^{12}$ Hz for physisorbed atoms and 10$^{13}$ Hz
for chemisorbed atoms  \cite{Hornekaer:2006I} and $T'$ is the `temperature' of the atom that is involved.

\begin{figure}
\begin{center}
\includegraphics{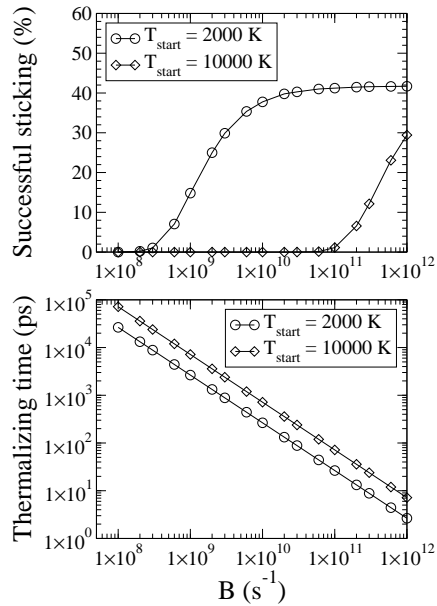}
\end{center}
\caption{Influence of the $B$ parameter on the sticking fraction (top) and the relaxation time (bottom) of the hot atom. For detailed explanation see text. }
\label{therma}
\end{figure}

In the experiments that we aim to reproduce, the atoms arrive at the surface at normal incidence and at a
temperature around 2000 K which is much higher than the surface temperature. Furthermore, the atoms will gain
energy due to the high binding energy if they chemisorb. The atoms will not be thermalized instantaneously, but
will most likely gradually lose their energy to the substrate. The dissipation of the excess energy into the
substrate is expected to be exponential \cite{Shalashilin:1998}. To reduce the computation complexity of the
model we emulate this exponential energy loss by a simpler expression and use the following function to describe
the temperature
\begin{equation}
T'(t) = \max\left(T_{\rm s}, \frac{T_{\rm start}}{\left(1+B \left(t-t_{\rm a}\right)\right)^2}\right) \label{T'}
\end{equation}
where $T_{\rm s}$ and $T_{\rm start}$ are the surface and starting temperature respectively, $t_{\rm a}$ is the
time at which the atom has adsorbed on the surface. The sensitivity of the model to the exact functional form was
checked and found to be small. For most cases we use $T_{\rm start} = 2000$ K, but also higher values of 7000 K
and  10,000 K are considered, matching the binding energy of a monomer. The parameter $B$ can be chosen freely.
Fig.~\ref{therma} studies the effect of this parameter. It displays the result of series of 10,000 simulations of
one deposition event for a particular value of B. The top panel indicates the percentage of deposition attempts
that resulted in sticking. The sticking fraction slowly approaches the monomer sticking barrier probability of 41
\% for increasing $B$. A value of $B$ above $\sim 10^{9}$ s$^{-1}$ for $T_{\rm start} = 2000$ K is needed to get
an initial sticking co-efficient in agreement with the experimental findings \cite{Zecho:2002}. Finally, the
bottom panel gives the relaxation time. For comparison Shalashilin and Jackson \cite{Shalashilin:1998} found for
a hydrogen atom on a Cu(111) surface that the thermal relaxation time is around 4 ps. This corresponds to a $B$
of $8\times 10^{10}$ s$^{-1}$ for $T_{\rm start} = 2000$ K and  $2\times 10^{11}$ s$^{-1}$ for $T_{\rm start} =
10,000$ K.

In the brief moment after a deposition, the rates in Eq.~\ref{R} are time dependent due to the decreasing
temperature and Eq.~\ref{t^i} cannot be used. Instead we use the method by Jansen et al.~\cite{Jansen:1995} to
determine $t^i$. This method makes use of
\begin{equation}
-\ln\left(X\right) = \int_t^{t^i}R_{\rm dif}^i(t) {\rm d}t + \int_t^{t^i}R_{\rm des}^i(t) {\rm d}t +\int_t^{t^i}R_{\rm
rec}^i(t){\rm d}t,
\label{integral}
\end{equation}
 that transforms to Eq.~\ref{t^i} if all rates are time dependent. To obtain $t_i$ the expression has to be solved. Using
\begin{eqnarray}
\Omega(t^i) &=& \int_{t}^{t^i} \nu \exp\left( -\frac{E(1+B(t-t_{\rm a}))^2}{kT_{\rm start}}\right) {\rm d} t\\
& =& \frac{\nu}{2B} \sqrt{\frac{\pi T_{\rm start}}{E}} {\rm erf}\left((1+B(t-t_{\rm a})) \sqrt{\frac{E}{T_{\rm
start}}}\right)
\end{eqnarray}
with ${\rm erf}$ the error function, Eq.~\ref{integral} becomes
\begin{equation}
-\ln\left(X\right) = \Omega_{\rm dif}(t^i) + \Omega_{\rm des}(t^i)  + \Omega_{\rm
rec}(t^i).
\end{equation}
This can be solved numerically to $t_i$ using the Newton-Raphson method \cite{NumRec}, since both $\Omega$ and
$\frac{{\rm d}\Omega}{ {\rm d}t} $ decrease monotonically. Notice that different functions for $T'$ will result
in a different expression for $\Omega(t^i)$. As the order in $t$ increases, solving $\Omega(t^i)$ becomes more
computationally expensive.

\section{Included reactions and energy barriers}
The previous section described the general Monte Carlo algorithm. For all processes energy barriers are needed to
determine the corresponding transition probabilities. These barriers are taken from independent DFT calculations
\cite{Sljivancanin:2007} of the binding energies of the different configurations that are considered, sticking
trajectories and diffusion trajectories. Since many possible configurations are formed during the simulations,
especially for high coverages, including all these different possibilities explicitly would make the Monte Carlo
program very slow and would require a huge set of barriers that all have to be calculated independently.  To
overcome this problem a number of simplifications are introduced resulting in the following sets of energy
barriers:

\subsection{Physisorbed atoms}
For the physisorption binding energy, the value of $\sim$ 40 meV is used based on results from selective
adsorption experiments \cite{Ghio:1980}. For diffusion an activation energy of 4 meV taken from
\cite{Bonfanti:2007} was used. We will come back to this diffusion rate at the discussion of mechanism IV. For
physisorbed atoms $T'(t) = T_{\rm s}$ was used.

\subsection{Sticking}
Numerous different barriers for sticking of an isolated hydrogen atom (a monomer) into the chemisorption site on
the graphite surface have been given. Jeloaica and Sidis found a barrier of $\sim$0.2 eV using the coronene
molecule as a model of a graphite surface \cite{Jeloaica:1999}. Sha et al. obtained a barrier slightly above 0.2
eV using a slab super cell with 4 layers each containing 8 carbon atoms \cite{Sha:2002,Sha:2002a}. Hornek{\ae}r
et al.~\cite{Hornekaer:2006II} found a barrier for chemisorption into a monomer of 0.15 eV using a single layer
super cell containing 32 carbon atoms. The influence of adding a second carbon layer was investigated and found
to be negligible. Using the same model surface a sticking barrier of 0.1 eV into the ortho-dimer and 0 eV into
the para-dimer configurations were found. All possibilities for sticking into a trimer state, starting from the
para-dimer, were seen to have non-zero sticking barriers between 0.1 and 0.15 eV. Adding a fourth atom resulting
in a triple para-dimer configuration again did not exhibit a barrier \cite{Hornekaer:2006II}. Barrierless
sticking into the para-dimer state was also found by Rougeau et al. \cite{Rougeau:2006}.

Even though there is a clear dependence on the local configuration of the impact site, we decided only to include
variations in sticking barriers for dimers in the simulation. The program determines if the incoming atoms can
form a dimer ignoring the larger configuration it might be part of and determines the barrier for sticking
accordingly. This assumption will cause deviations between simulations and experimental results at high coverage,
since it overestimates the formation of trimer configurations that contain para-dimers. No sticking barrier was
used for physisorption of H atoms. The barriers for sticking used in the simulation are summarized in Table
\ref{Echem}.

Section \ref{Sense} tests the different mechanisms to their sensitivity to various input parameters. The monomer sticking barrier is one of them.

\begin{table}
\caption{The sticking barrier for different configurations}
\label{Echem}
\begin{tabular}{lr@{.}l}
\hline
\hline
Configuration & \multicolumn{2}{c}{$E_{\rm stick}$ [meV]}\\
\hline
\hline
para-dimer    &  0&0  \\
ortho-dimer   &  0&1  \\
\hline
other chemisorption site       &  0&15\\
\hline
physisorption        &  0&0\\
\hline \hline

\end{tabular}
\end{table}

\begin{table}
\caption{The total binding energy for different configurations}
\label{Ebin}
\begin{tabular}{lr@{.}l}
\hline
\hline
Configuration$^1$   & \multicolumn{2}{c}{$E_{\rm bind}$ [meV]}\\
\hline
\hline
monomer           & -0&8 \\
\hline
para-dimer (G)    & -2&9 \\
ortho-dimer (A)   & -2&8 \\
meta-dimer (E)    & -0&8 \\
\hline
trimer (A, G)     & -3&9 \\
trimer (G, I)     & -3&7 \\
trimer (D, G)     & -3&5 \\
trimer (B, G)     & -3&5 \\
trimer (F, G)     & -3&3 \\
\hline
tetramer (A, G, H)& -5&9 \\
tetramer (A, B, E)& -5&9 \\
tetramer (A, D, G)& -5&7 \\
tetramer (A, B, D)& -5&5 \\
tetramer (B, E, G)& -5&1 \\
tetramer (A, B, C)& -5&0 \\
tetramer (A, B, G)& -4&8 \\
tetramer (A, B, F)& -4&5 \\
\hline
\hline
\end{tabular}

{\footnotesize $^1$ See Fig.~\ref{nummers}.}
\end{table}

\begin{figure}
\begin{center}
\includegraphics{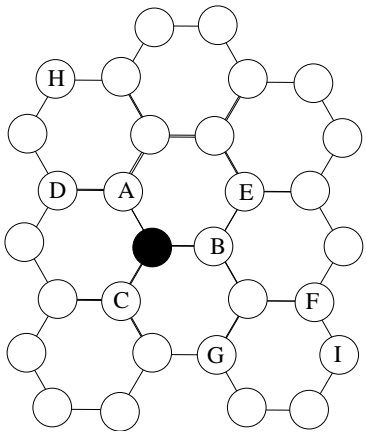}
\end{center}
\caption{Schematic guide to obtain the configurations used in Table \ref{Ebin}. A configuration is made up from hydrogen atoms positioned on top of the black carbon atom and the atoms indicated by the characters given in Table \ref{Ebin}. In this way, an ortho-dimer is represented by the character (A). }
\label{nummers}
\end{figure}

\subsection{Binding energies}
The binding energies for different configurations of chemisorbed H atoms are displayed in table II. The different
configurations are displayed in Fig.~\ref{nummers}. These values are based on DFT calculations reported in
\cite{Hornekaer:2006II,Sljivancanin:2007}.

\subsection{Diffusion}
\begin{figure}
\begin{center}
\includegraphics{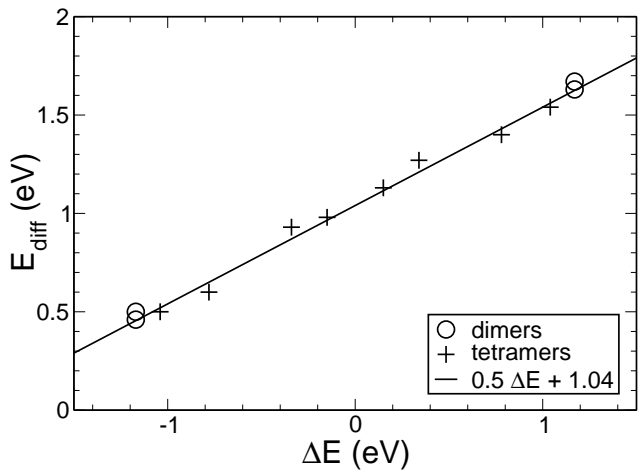}
\end{center}
\caption{The diffusion barrier as a function of the energy difference between the initial and final configuration. The diffusion barriers between dimer and the tetramer configurations follow the same linear dependences.}
\label{Eb}
\end{figure}
For several dimer and tetramer configurations the individual binding energies and the transition barriers between
some of the configurations were determined via DFT calculations. The binding energy of the individual dimer and
tetramer configurations are given in Table \ref{Ebin}. The tetramer binding values and diffusion barriers are
taken from \cite{Sljivancanin:2007}. Figure \ref{Eb} plots the diffusion barrier as a function of the energy
difference of the configurations. As the figure clearly shows there is a linear relation between the two
quantities and dimers and tetramers follow the same relation. A least-squares fit resulted in
\begin{equation}
E_{\rm diff} = 0.5 \Delta E + 1.04 \textrm{ eV}.
\label{E_diff}
\end{equation}
In order to obey detailed balance, or microscopic reversibility, transition probabilities for diffusion should
fulfill
\begin{equation}
\frac{P_{ij}}{P_{ji}} = \exp\left(-\frac{\Delta E_{ij}}{kT}\right)
\end{equation}
Using Eq.~\ref{R}, it can be shown that the empirically found relation, Eq.~\ref{E_diff}, follows this requirement.

\begin{figure*}
\includegraphics{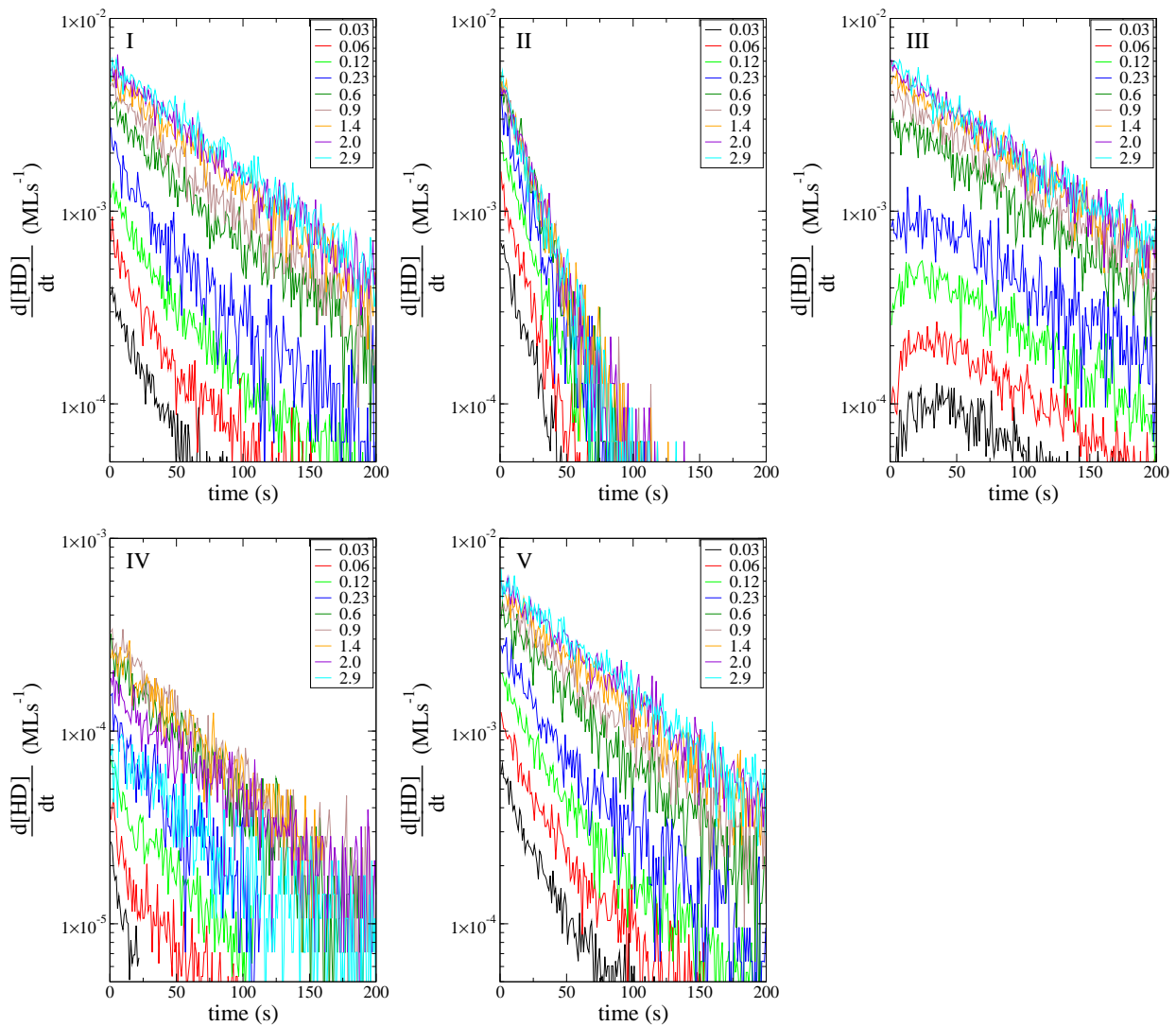}
\caption{The amount of HD molecules per time leaving the surface in ML/s as a function of time for different
pre-exposures of D. At $t=0$ the hydrogen beam is switched on. For (I) mechanism I: Direct Eley-Rideal ($E_{\rm
ER, mono} = E_{\rm ER, dimer} = 9$ meV), (II) mechanism II: Direct Eley-Rideal with steering ($E_{\rm
ER, mono} = E_{\rm ER, dimer} = 9$ meV, $s = 1$), (III) mechanism III: Dimer Eley-Rideal ($E_{\rm ER, mono} = \infty;E_{\rm
ER, dimer} = 0$ meV), (IV) mechanism IV: Dimer mediated Eley-Rideal ($B = 1\times 10^{9}$ s$^{-1}$), and (V) mechanism V: Direct Eley-Rideal with fast diffusion of physisorbed atoms ($E_{\rm
ER, mono} = E_{\rm ER, dimer} = 9$ meV, $R_{\rm dif} = 5 \times 10^{13}$ s$^{-1}$).}
\label{dHDdt}
\end{figure*}

\subsection{Desorption}
Finding a general expression for the desorption energy of the individual atoms from a configuration is less
straightforward. The binding energies are determined for the configuration and not for the individual atoms.
Again calculating all possible desorption pathways would not be feasible. We base the desorption on the total
binding energy of the configuration of $n$ atoms, $E_{\rm bind}$. The desorption energy is then
\begin{equation}
E_{\rm des} = \frac{E_{\rm bind}}{n} + E_{\rm stick}
\end{equation}
where $E_{\rm stick}$ is the barrier for sticking in the same position.  $E_{\rm stick}$ and $E_{\rm bind}$ can be found in Tables \ref{Echem} and \ref{Ebin}.

\subsection{Thermally activated H$_2$ formation}
Formation of H$_2$ via reaction between two chemisorbed hydrogen atoms to gaseous molecular hydrogen occurs from
the dimer states and has to overcome barriers of 2.49 eV for the ortho-dimer and 1.4 eV for the para-dimer state
\cite{Hornekaer:2006I}. Reaction barriers from the isolated dimer states are used for all configurations. This
approximation will again lead to inaccuracies for simulations at high coverage.

\subsection{Abstraction}

Five different Eley-Rideal abstraction mechanisms are considered in the model:
\begin{enumerate}
\item direct Eley-Rideal with all hydrogen atoms regardless of their local configuration.
\item  a simple version of Eley-Rideal with steering where a direct Eley-Rideal reaction is allowed not just for
H atoms impinging on an already occupied site but also for H atoms impinging on adjacent sites.
\item direct Eley-Rideal with only hydrogen atoms part of a para-dimer.
\item a dimer mediated reaction where the incoming atom is first adsorbed into a dimer configuration and, before
thermalizing to the substrate temperature, forms H$_2$ and desorbs.
\item direct Eley-Rideal together with a high diffusion rate of the atoms in the physisorption state
\end{enumerate}

The influence of the different Eley-Rideal mechanisms can be controlled by using different barriers for the
direct Eley-Rideal reaction and different values of the thermalization parameter B. Furthermore, a simple version
of steering is implemented with an adjustable parameter $s$ as will be described below. The values of the parameters used for simulating the different mechanisms
are given in Table \ref{Abs}.

The height of the abstraction barrier is somewhat uncertain due to the limitations in the accuracy of DFT
calculations. Morisset et al. \cite{Morisset:2004} found a barrier for direct abstraction of a monomer just below
10 meV. Others also found low barriers to abstraction \cite{Sha:2002a,Martinazzo:2006}. Following Morisset et al.
we employ a barrier of 9 meV but later investigate the effect of changing the value of the barrier. The same
group also found that the abstraction reaction with one of the hydrogen atoms in a para-dimer can proceed without
barrier \cite{Bachellerie:2007}.

\begin{table*}
\caption{Model parameters for the different abstraction mechanisms} \label{Abs}
\begin{tabular}{l|l|l|l|l|l|l}
\hline
Mechanism & {$E_{\rm ER,mono}$ [meV]}     & {$E_{\rm ER,dimer}$ [meV]}      & {$B$ [$s^{-1}$]} & {$T_{\rm start}$ [K]} & {$s$} & $R_{\rm dif}$[$s^{-1}]$\\
\hline
I (Fig.~\ref{dHDdt}-I)     &        9 &        9 & $10^{12}$ & 2000 & 0  & $7.2 \times 10^{12}$ \\
II (Fig.~\ref{dHDdt}-II)   &        9 &        9 & $10^{12}$ & 2000 & 1  & $7.2 \times 10^{12}$\\
III (Fig.~\ref{dHDdt}-III) & $\infty$ &        0 & $10^{12}$ & 2000 & 0  & $7.2 \times 10^{12}$\\
IV (Fig.~\ref{dHDdt}-IV)   & $\infty$ & $\infty$ & $10^9$    & 2000 & 0  & $7.2 \times 10^{12}$\\
V (Fig.~\ref{dHDdt}-V)     &        9 &        9 & $10^{12}$ & 2000 & 0  & $5.0 \times 10^{13}$\\
\hline

\end{tabular}
\end{table*}

\section{Comparison with experiments}
The Monte Carlo simulation results are compared with the experimental data on the Eley-Rideal reaction presented
by Zecho et al. \cite{Zecho:2002}. These experiments consist of two phases. First, a graphite substrate was
exposed to a normal incidence deuterium beam with a flux of $3.8 \times 10^{15}$ atoms cm$^{-2}$s$^{-1}$ at 150
K. The exposure time was varied to give a range of initial coverages. The temperature of the atom beam was 2000
K. During the second phase the pre-exposed substrate was exposed to a normal incidence hydrogen beam and the
formed HD molecules were measured using a mass spectrometer. The cross section of the deuterium abstraction with
hydrogen, $\sigma$, was determined from these spectra using
\begin{equation}
\frac{{\rm d[HD]_{\rm g}}}{{\rm d}t} = \sigma \Phi {\rm[D_{\rm ad,0}]} \exp\left(-\sigma \Phi t\right),\label{dHD/dt}
\end{equation}
with $\Phi$ the H flux and ${\rm[D_{\rm ad,0}]}$ the initial D coverage. For the derivation of this equation, we
refer to \cite{Zecho:2002}. The expression is obtained assuming only direct Eley-Rideal. Since the reaction
mechanism is implicitly included in the cross section, it is hard to directly interpret the results obtained in
this way, but we will use the method to compare the experimental data with our simulation results.

The hydrogen adsorbate configurations found in the Monte Carlo simulations are also compared to the adsorbate
configurations observed in STM experiments. In particular the fraction of hydrogen atoms in dimer configurations
or larger clusters in simulation and experiment are compared.

\section{Results}
Several simulation runs were performed using a similar set of pre-exposures as used by Zecho et al.
\cite{Zecho:2002}. The H and D fluxes were $4.8 \times 10^{13}$ cm$^{-2}$s$^{-1}$ at a temperature of 150 K. $250
\times 125$, $500 \times 250$, and $750 \times 500$ chemisorption sites were used depending on the initial D
coverage. This corresponds to an array of $250 \times 250$, $500 \times 500$, and $750 \times 750$ to accommodate
the extra physisorption sites. If the noise was still considerable for arrays of $750 \times 500$ multiple runs
with different random seeds were made. In order to test to what extent the five proposed abstraction mechanisms
contribute to the Eley-Rideal abstraction process comparisons to the experimental results, with different
parameter sets (see Table \ref{Abs}) were performed.

Figure \ref{dHDdt} shows ${\rm d[HD]_{\rm g}}/{{\rm d}t}$ for the five different abstraction mechanisms. Panel (I)
only includes the direct Eley-Rideal with a barrier of 9 meV independent of the configuration on the surface,
panel (II) investigates the influence of steering, panel (III) only allows an Eley-Rideal reaction, with a zero barrier, if the surface atom is part of a para-dimer
configuration, panel (IV) uses the slower thermalization with $B = 1\times 10^{8}$ s$^{-1}$ to test the dimer
mediated mechanism, and panel (V) uses very fast diffusion. As ${\rm d[HD]_{\rm g}}/{{\rm
d}t}$ is plotted on a logarithmic scale, the curves should be linear according to Eq.~\ref{dHD/dt} with the slope
the cross section times the flux. The flux has the same value throughout the simulations.

The key results of the simulations are summarized in Table \ref{Sumtable} and compared to the experimental
findings. Each entry consists of nine individual simulations. The top twelve rows represent the runs that allow only
a single mechanism. The second section summarizes simulation series of combinations of mechanisms and the third
part gives the results with a different sticking barrier and will be discussed in Section \ref{Sense}. The final
entries give the experimental results for comparison. The difference in cross section between high and low
coverage, the saturation coverage, and the linearity of the initial signal in initial coverage are indicators of
the agreement with the experiment. The latter linearity is measured by the Pearson correlation coefficient that
gives one for perfect correlation and zero for no correlation. For a better comparison between the different
simulation runs all cross sections at low coverage are given for $\sim0.013$ ML which is generally achieved
after an exposure of 0.03 ML (the lowest considered pre-exposure). Simulations including mechanism IV (dimer mediated reaction) need a
longer exposure to reach 0.013 ML whereas simulations with mechanism V (fast diffusion) reach it faster than 0.03 ML. The cross
sections at high coverage are taken at the highest considered pre-exposure of 2.9 ML. The tenth column gives the corresponding coverages. In almost all cases the saturation coverage has been reached at this point. The last column of the
table indicates the fraction of atoms that is part of a dimer configuration or larger cluster at a coverage of
0.5 \% at conditions similar to the ones used by Hornek{\ae}r at al.~\cite{Hornekaer:2006I}.

\subsubsection*{Mechanism I: Direct Eley-Rideal}
\begin{figure}
\includegraphics{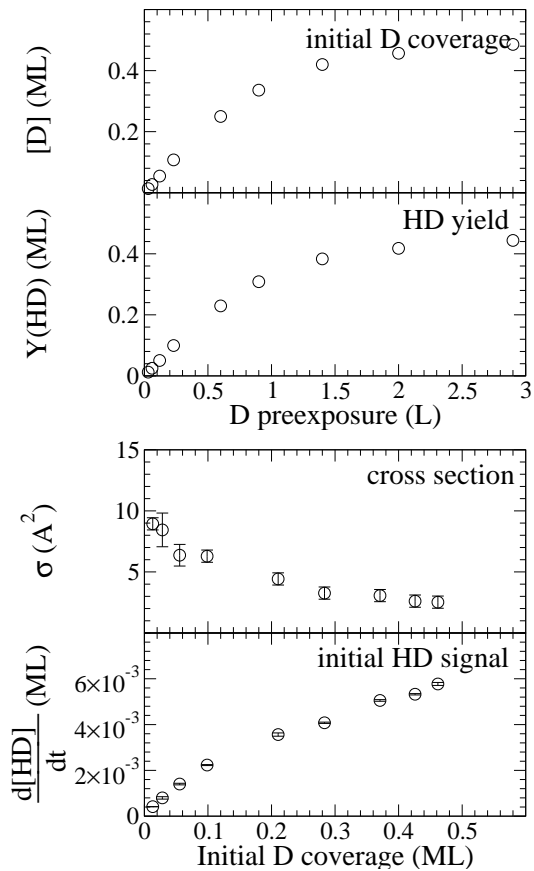}
\caption{Analysis of Figure \ref{dHDdt}-I ($E_{\rm ER, mono} = E_{\rm ER, dimer} = 9$ meV). (a) Initial coverage and yield as a function the D pre-exposure. (b) The cross section and initial signal versus the initial D coverage.}
\label{cs_1}
\end{figure}

The five plots in Fig. \ref{dHDdt} all show a very different behavior. Panel (I), which tests the direct Eley-Rideal
mechanism, shows a linear relation between $\ln\left({\rm d[HD]_{\rm g}}/{{\rm d}t}\right)$ and
the deposition time over the whole time range. 
The slope appears to have some dependence of the initial D coverage. This would result in a
cross section that is dependent of initial coverage. Figure \ref{cs_1} plots, among other quantities, the cross
section which indeed shows some coverage dependence. It is however much less than seen in the experiments \cite{Zecho:2002}. The cross section and the value of the initial signal
is obtained by fitting the curves shown in Figure \ref{dHDdt}-I to Eq.~\ref{dHD/dt}. Since for some parameter
choices only the first section is linear, only these points are included in the fit. The number of points
included in the fit can have a strong effect on the final result, especially if the linear part is very short.
This is reflected by the error bars. Various runs with different initial seed showed that this error should at least be 0.5 \AA$^2$. The initial signal depends linearly on the deuterium coverage, which agrees
with Figure 3c in \cite{Zecho:2002}.  Figure \ref{cs_1} also displays the total HD yield. Since this
value is comparable with the initial D coverage it is clear that only a negligible fraction of D atoms remains on
the surface after hydrogen exposure. This is in good agreement with the experimental findings \cite{Zecho:2002}. The dimer fraction is very low in contrast with the experiments as can be seen in the first row in Table \ref{Sumtable}.

\subsubsection*{Mechanism II: Eley-Rideal with steering}
\begin{figure}
\includegraphics{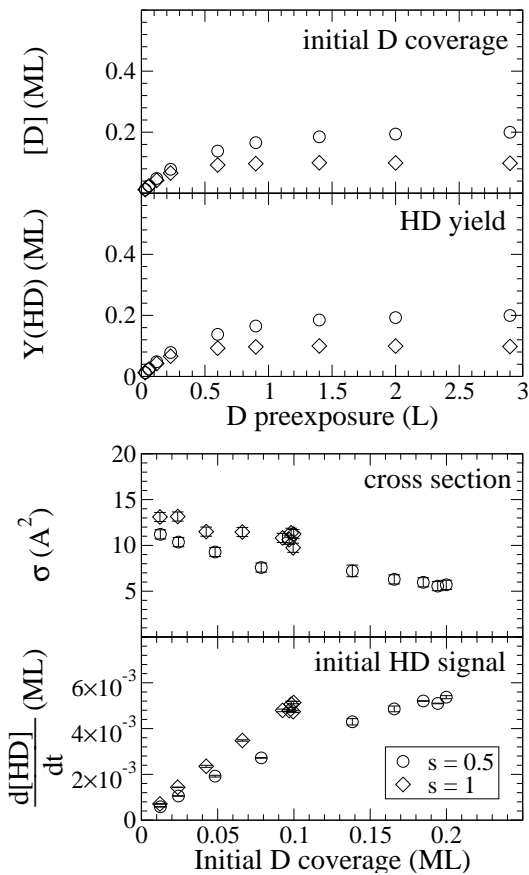}
\caption{Analysis of Figure \ref{dHDdt}-II ($E_{\rm
ER, mono} = E_{\rm ER, dimer} = 9$ meV and steering). (a) Initial coverage and yield as a
function the D pre-exposure. (b) The cross section and initial signal versus the initial D coverage.}
\label{cs_4}
\end{figure}

 A simple version of Eley-Rideal abstraction with steering is implemented by allowing atoms that land on
empty sites close to chemisorbed hydrogen atoms to attempt a reaction with these atoms. For the
reaction barrier the same $E_{\rm ER,mono}$ is used as for the direct hit. Eley-Rideal via a direct hit
(mechanism I) is also allowed. If an atom lands on an empty site, the possibility of steering is considered by
comparing a random number between 0 and 1 against parameter $s$. If the random number is smaller than $s$, then
steering is activated. In the case of steering, the three neighbouring sites are checked for reacting species. If
two or more different atoms are found, another random number determines with which of the atoms the incoming H
atom will undergo the reaction attempt. Simulations for $s=1$ and 0.5 were performed. The results of these two
simulation runs are summarized in Fig.~\ref{cs_4} and the rows indicated by II in Table \ref{Sumtable} and the $s=1$ simulation run is shown in Figure \ref{dHDdt}-II.
Fig.~\ref{cs_4} shows the initial coverage and yield after 200 seconds of H exposure as function of the initial D
pre-exposure. Notice that the yield is given in ML in these graphs and not as a percentage. It further gives the
cross section and initial signal as a function of the initial coverage. For $s=1$, a reasonably high cross
section at low coverage is obtained ($14.7 \pm 1.4$ \AA$^2$), but the saturation coverage is very low (0.10 ML)
and also the cross section at this coverage is too high. The series which simulates steering with $s=0.5$ gives
somewhat lower cross section resulting in better agreement with the experimental findings at high coverage and
too low a cross section at low coverage. Furthermore, an almost linear decrease in cross section is found as a
function of coverage in disagreement with the experimental findings. Again the dimer ratio is too low.

\subsubsection*{Mechanism III: Dimer Eley-Rideal}
The curves in panel (III) are not linear and show a maximum at later times for the low coverages. Since this panel
presents the results with the dimer Eley-Rideal mechanism, this maximum indicates the time at which a maximum of
D containing dimers is formed. This behavior contradicts the experiments \cite{Zecho:2002} where the [HD] signal decreases over time and is therefore not
likely to be the primary mechanism involved in the abstraction.

\begin{figure}
\includegraphics{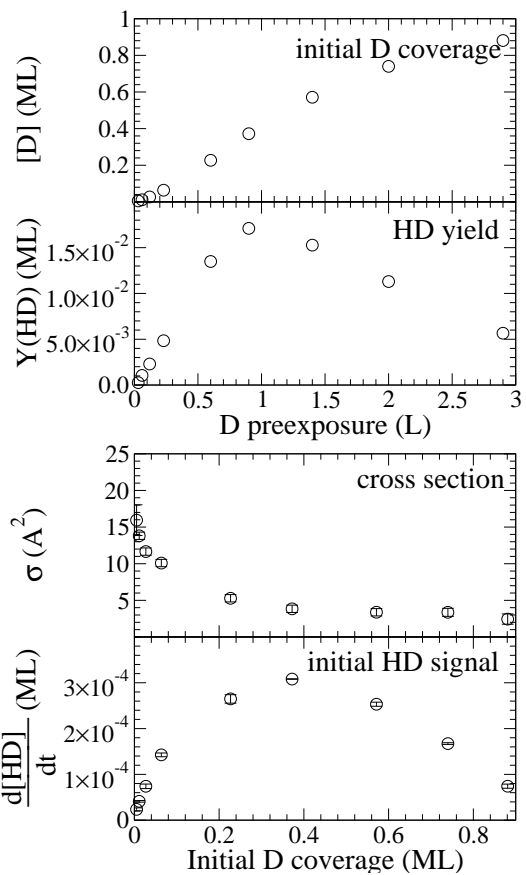}
\caption{Analysis of Figure \ref{dHDdt}-IV ($B = 1 \times 10^8$ s$^{-1}$). (a) Initial coverage and yield as a
function the D pre-exposure. (b) The cross section and initial signal versus the initial D coverage.}
\label{cs_3}
\end{figure}

\subsubsection*{Mechanism IV: Dimer mediated abstraction}
Panel (IV) presents the results for the dimer mediated mechanism. Here a clear linear dependence can be observed
for very early times with a decreasing slope for increasing initial coverage. To obtain this graph large arrays
upto $1500 \times 3000$ sites were used due to the low sticking rate at low coverage. Notice that the y-axis
range of this panel is different from the others. Figure \ref{cs_3} studies this set of simulations more closely.
The two upper panels show that not all D is converted into HD, but  that a substantial amount of deuterium still
resides at the surface after the 200 seconds exposure.  This is in contrast with the experimental finding that
all deuterium is abstracted after a 3 ML dose. Figure \ref{cs_3} also shows the cross section determined from
Fig.~\ref{dHDdt}-II. Again the bars indicate the uncertainties in obtaining the slope. Since the linearity is much
less than for mechanism I, the error bars are larger than in Fig.~\ref{cs_1}. In contrast with this latter
figure, Figure \ref{cs_3} shows a strong dependence of the cross section on the initial deuterium coverage. This
corresponds with the trends observed by Zecho et al. \cite{Zecho:2002}. The values are also in the correct range,
although the cross section is too high at low coverage and too low for high coverage. Both the cross section and
the initial signal indicate that the dimer mediated mechanism becomes very inefficient for high coverages. The
initial signal even decreases for increasing surface coverage indicating that at high coverage it is hard to make
dimers due to the decreasing number of available sites. As the coverage increases, either by H or by D atoms, the
mechanism becomes less efficient and the deuterium atoms will not be abstracted. This also results in a very high
saturation coverage.

The dimer fraction is higher as compared to the other mechanisms, although still much too low. The increase in
dimer fraction is due to the fact that many of the chemisorbed atoms desorb due to the initial energy. Since
atoms in dimer position are less likely to desorb because of their higher binding energy, this results in an
elevated dimer ratio.

\begin{figure}
\includegraphics{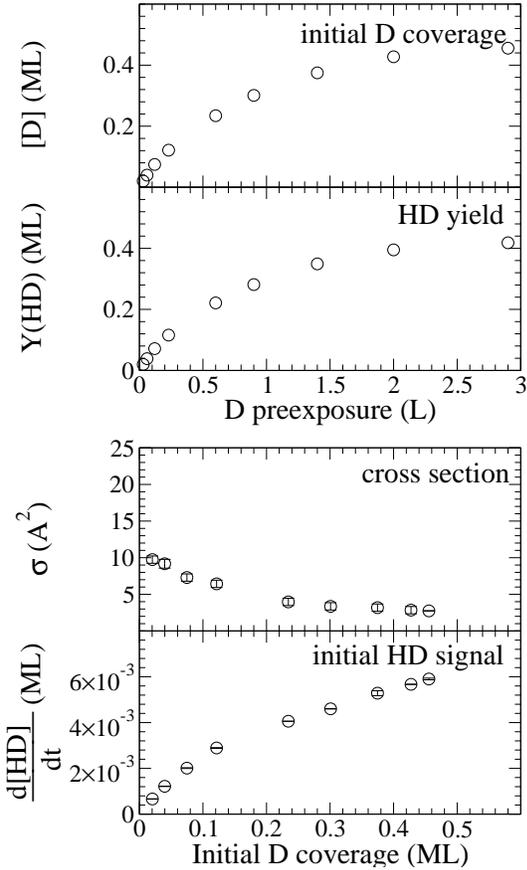}
\caption{Simulations including mechanism V ($R_{\rm dif} = 5.0
\times 10^{13}$ s$^{-1}$). (a) Initial coverage and yield as a function the D pre-exposure. (b) The cross section
and initial signal versus the initial D coverage.}
\label{cs_5}
\end{figure}

\subsubsection*{Mechanism V: Fast diffusion}
As discussed above, the barriers to diffusion in the chemisorbed state are quite high making this process
essentially negligible at least at low coverage. Diffusion in the physisorbed state is, however, a completely
different matter. If a simple expression for an activated process is assumed, then an atom thermalized to a
surface temperature of 150 K will make around 20 hops, which corresponds to a distance of 5 {\AA} before
desorbing. However, as Bonfanti et al.~\cite{Bonfanti:2007} pointed out diffusion is not an activated process,
since the activation barrier of 4 meV is negligible at these temperatures. This asks for a different estimation
of the moving rate. As a first approach we took the diffusion coefficient obtained by Bonfanti et
al.~\cite{Bonfanti:2007}. This results in a moving rate of $1.3 \times 10^{13}$ s$^{-1}$. The upper limit for
diffusion will be free movement in two dimensions according to the average gas phase velocity. This results in a
diffusion rate of $5.0 \times 10^{13}$ s$^{-1}$ at 150 K. We will perform simulations using both rates. The
results for the highest rate and $E_{\rm ER,mono} = 9$ meV are shown in Figures \ref{dHDdt}-V and \ref{cs_5}.
Results of other simulation series with different combinations of $E_{\rm ER,mono}$ and the diffusion rate are
summarized in Table \ref{Sumtable}. A clear increase in the dimer fraction can be observed for increasing
diffusion rate, although still slightly too low compared with experiments. The cross section curve as a function
of the coverage has the correct dependence although it is still not high enough at low coverage. Furthermore  the
initial signal as a function of coverage does not have a linear dependence but exhibits a kink around 10 \%. At
this coverage the number of trimers starts declining in favour of the number of tetramers. Most of the
configurations (60 \%) are now a configuration of three or higher. This non-linearity could well be an artifact
of the simplified treatment of reaction and sticking for these configurations which makes the model inaccurate at 
high coverage as discussed above.

\subsubsection*{Combined mechanisms}

The experiments cannot be explained by one of the five mechanisms separately. Mechanism I is not efficient enough
at low coverages, mechanism II leads to very low saturation coverages, mechanism III shows a maximum at $t>0$,
mechanism IV is not efficient at high coverages. Moreover, all these four mechanisms show very low dimer fractions
compared with  experiments. Mechanism V has an increased number of dimers and a higher cross section, but again
this is not high enough to match the experimental observations. We therefore investigate different combinations
of the five mechanisms.

\begin{figure}
\includegraphics{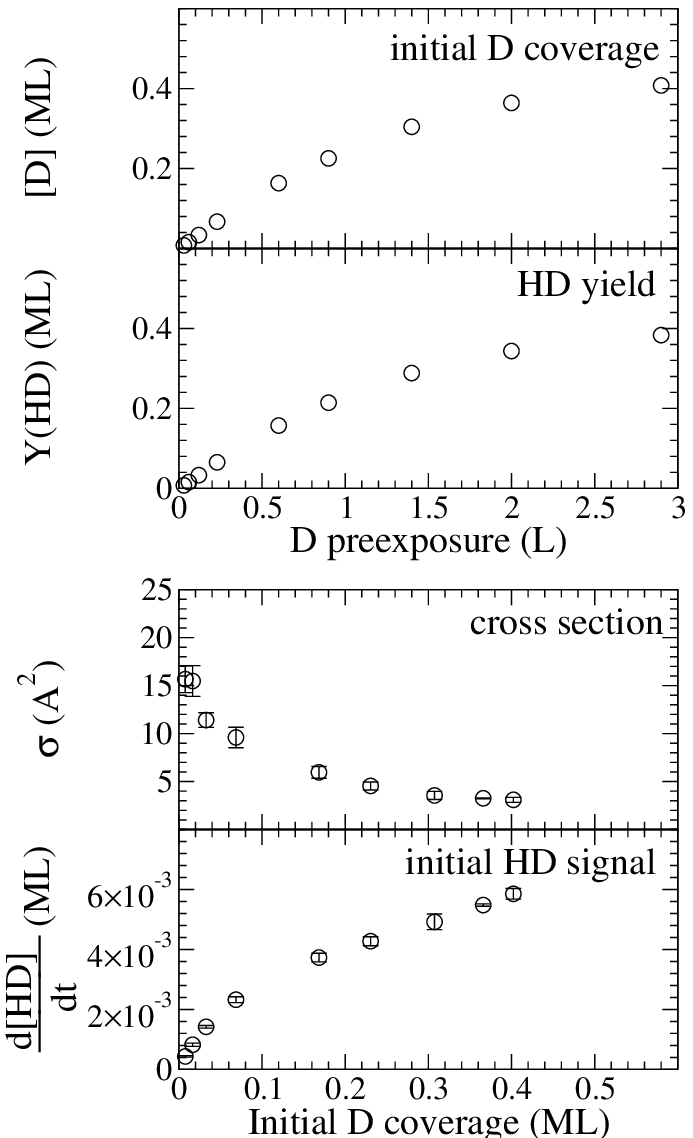}
\caption{Simulations including mechanism IV ($B = 1
\times 10^9$ s$^{-1}$) and mechanism V ($E_{\rm ER, mono} = E_{\rm ER, dimer}  = 9$ meV and $R_{\rm dif} = 5.0\times 10^{13}$ s$^{-1}$). (a) Initial coverage and yield as a function the D pre-exposure. (b) The cross section
and initial signal versus the initial D coverage.}
\label{cs_3+5}
\end{figure}

Again the results are summarized in Table \ref{Sumtable}. The table clearly shows that a combination of
mechanisms IV and V gives the best results, both in terms of dimer ratios and cross section. The best agreement is
obtained for the high diffusion rate and $E_{\rm ER,mono} = 9$ meV (in bold face). The detailed analysis of these simulations is
given in Figure \ref{cs_3+5}.  The cross section has the correct coverage dependence, both in trend and in value.
Also the saturation coverage of $0.4 \pm 0.2$ ML found experimentally is reproduced. The major discrepancy is again
the initial HD signal which does not dependent linearly on the initial cross section, but as mentioned earlier
this can be an artifact due to the simplified assumptions made for sticking and reaction for complex
configurations, which make high coverage simulations inaccurate. The dimer ratio is still below the experimental
values, but much higher than all previous series. The dimer ratio includes both ortho and para dimers, but the
main contribution is from the para dimers (40 times more at 0.1 \%), since these sites have no barrier for
sticking. The residence time of the physisorbed atoms at ortho sites is not long enough for the atoms to easily
enter the chemisorbed state. As the dimers have been observed in more equal ratios, the rate for crossing this
barrier is probably higher, possibly due to a lower barrier or higher pre-exponential factor than used in the
present model or due to a longer residence time at the ortho site because of the irregularity in the lattice
caused by the first hydrogen atom. The reason that this combination of mechanisms give higher cross section at
low coverage is because more reaction events go via a dimer state. In this state an atoms gains some energy
due to the strong binding energy and this extra binding energy is used to overcome the barrier for reaction.

Other combinations show less agreement with the experiments on at least two points of comparison.

\begin{table*}
\caption{Simulation results for different parameter settings}
\label{Sumtable}
\begin{tabular}{l|r|l|r|r|c|r@{.}l@{ $\pm$ }r@{.}l|r@{.}l@{ $\pm$ }r@{.}l|r@{.}l|r@{.}l|c}
\hline
\hline
Mechanism & \multicolumn{1}{|c}{$E_{\rm ER,mono}$} & \multicolumn{1}{|c|}{$B$} & $T_{\rm start}$ & \multicolumn{1}{|c|}{$R_{dif}$} & steering & \multicolumn{4}{c}{Cross section} & \multicolumn{4}{|c}{Cross section}  & \multicolumn{2}{|c}{Correlation} & \multicolumn{2}{|c|}{[D]$_0$ (2.9 L)} & Dimers\\
 & \multicolumn{1}{|c}{(meV)} & \multicolumn{1}{|c|}{(s$^{-1})$} & $(K)$ & \multicolumn{1}{|c|}{(s$^{-1})$} & $s$ & \multicolumn{4}{c}{low [D]$_0$ (\AA$^2$)} & \multicolumn{4}{|c}{high [D]$_0$ (\AA$^2$)} & \multicolumn{2}{|c}{coefficient} & \multicolumn{2}{|c|}{(ML)}  & \%\\
\hline
\hline
I (Fig.~\ref{dHDdt}-I) &  9 & $2 \cdot 10^{12}$&  2000 & $7.2\cdot 10^{11}$ & 0 &  8&9 & 0&7 & 2&5 & 0&5 & 0&987 & 0&46 & 13\\
                      & 18 & $2 \cdot 10^{12}$&  2000 & $7.2\cdot 10^{11}$ & 0 &  7&5 & 0&5 & 2&6 & 0&5 & 0&995 & 0&48 & 13\\
                      &  0 & $2 \cdot 10^{12}$&  2000 & $7.2\cdot 10^{11}$ & 0 &  9&9 & 0&8 & 3&0 & 0&5 & 0&985 & 0&44 & 13\\
\hline
II (Fig.~\ref{dHDdt}-II)&  9 & $2 \cdot 10^{12}$&  2000 & $7.2\cdot 10^{11}$ & 1 & 13&1 & 1&1 &11&4 & 0&5 & 0&997 & 0&10 & 11\\
                      &  9 & $2 \cdot 10^{12}$&  2000 & $7.2\cdot 10^{11}$ &0.5& 11&2 & 0&7 & 5&7 & 0&5 & 0&993 & 0&20 & 12\\
\hline
III (Fig.~\ref{dHDdt}-III)&   & $2 \cdot 10^{12}$&  2000 & $7.2\cdot 10^{11}$ & 0 &  0&0 & 0&0 & 2&5 & 0&5 & 0&998 & 0&52 & 21\\
\hline
IV (Fig.~\ref{dHDdt}-IV)&    & $1 \cdot 10^{9}$ &  2000 & $7.2\cdot 10^{11}$ & 0 & 13&8 & 1&2 & 2&4 & 0&7 & 0&302 & 0&88 & 29\\
                      &    & $4 \cdot 10^{11}$& 10000 & $7.2\cdot 10^{11}$ & 0 &  6&4 & 1&0 & 2&2 & 0&5 & 0&856 & 0&54 & 72\\
\hline
V                     &  9 & $2 \cdot 10^{12}$&  2000 & $1.3\cdot 10^{13}$ & 0 & 11&8 & 2&0 & 2&7 & 0&5 & 0&977 & 0&46 & 32\\
                      & 18 & $2 \cdot 10^{12}$&  2000 & $1.3\cdot 10^{13}$ & 0 &  6&5 & 0&6 & 2&4 & 0&5 & 0&989 & 0&47 & 32\\
 (Fig.~\ref{dHDdt}-V) &  9 & $2 \cdot 10^{12}$&  2000 & $5.0\cdot 10^{13}$ & 0 & 10&1 & 0&5 & 2&8 & 0&5 & 0&984 & 0&46 & 52\\
                      & 18 & $2 \cdot 10^{12}$&  2000 & $5.0\cdot 10^{13}$ & 0 &  5&8 & 0&5 & 2&6 & 0&5 & 0&982 & 0&47 & 53\\
\hline \hline
I + III               &  9 & $2 \cdot 10^{12}$&  2000 & $7.2\cdot 10^{11}$ & 0 &  9&5 & 0&8 & 2&8 & 0&5 & 0&986 & 0&45 & 21\\
\hline
I + IV                &  9 & $1 \cdot 10^{9}$ &  2000 & $7.2\cdot 10^{11}$ & 0 & 10&0 & 1&0 & 3&1 & 0&5 & 0&984 & 0&42 & 30\\
\hline
II + IV               &  9 & $1 \cdot 10^{9}$ &  2000 & $7.2\cdot 10^{11}$ & 1 & 14&2 & 0&8 &11&0 & 0&5 & 0&994 & 0&08 & 27\\
\hline
III + V               &  9 & $2 \cdot 10^{12}$&  2000 & $5.0\cdot 10^{11}$ & 0 &  8&3 & 0&8 & 2&7 & 0&5 & 0&983 & 0&45 & 52\\
\hline
II + V                &  9 & $2 \cdot 10^{12}$&  2000 & $1.3\cdot 10^{13}$ & 1 & 14&3 & 0&9 &10&5 & 0&5 & 0&998 & 0&12 & 30\\
                      & 18 & $2 \cdot 10^{12}$&  2000 & $1.3\cdot 10^{13}$ & 1 & 11&9 & 0&5 & 9&1 & 0&5 & 0&996 & 0&13 & 30\\
                      &  9 & $2 \cdot 10^{12}$&  2000 & $5.0\cdot 10^{13}$ & 1 & 13&6 & 1&0 &10&4 & 0&5 & 0&997 & 0&13 & 50\\
                      & 18 & $2 \cdot 10^{12}$&  2000 & $5.0\cdot 10^{13}$ & 1 & 11&4 & 0&5 & 9&1 & 0&5 & 0&997 & 0&14 & 51\\
                      &  9 & $2 \cdot 10^{12}$&  2000 & $1.3\cdot 10^{13}$ &0.5& 10&9 & 0&7 & 6&6 & 0&5 & 0&986 & 0&20 & 33\\
                      & 18 & $2 \cdot 10^{12}$&  2000 & $1.3\cdot 10^{13}$ &0.5& 10&5 & 1&5 & 5&6 & 0&5 & 0&992 & 0&22 & 33\\
                      &  9 & $2 \cdot 10^{12}$&  2000 & $5.0\cdot 10^{13}$ &0.5& 10&9 & 0&5 & 5&7 & 0&5 & 0&987 & 0&21 & 52\\
                      & 18 & $2 \cdot 10^{12}$&  2000 & $5.0\cdot 10^{13}$ &0.5&  9&6 & 0&5 & 5&3 & 0&5 & 0&996 & 0&23 & 52\\
\hline
IV + V                &  9 & $1 \cdot 10^{9}$ &  2000 & $1.3\cdot 10^{13}$ & 0 & 12&8 & 0&7 & 3&0 & 0&5 & 0&971 & 0&40 & 55\\
                      & 18 & $1 \cdot 10^{9}$ &  2000 & $1.3\cdot 10^{13}$ & 0 & 11&9 & 1&2 & 2&7 & 0&5 & 0&987 & 0&43 & 53\\
\bf (Fig.~\ref{cs_3+5})  &\bf  9 & \boldmath$1 \cdot 10^{9}$ &\bf  2000 & \boldmath$5.0\cdot 10^{13}$ &\bf 0 &\bf 15&\bf6 &\bf 1&\bf5 &\bf 3&\bf2 &\bf 0&\bf5 &\bf 0&\bf980 & \bf0&\bf40 & \bf74\\
                      & 18 & $1 \cdot 10^{9}$ &  2000 & $5.0\cdot 10^{13}$ & 0 & 10&8 & 0&7 & 2&8 & 0&5 & 0&985 & 0&43 & 76\\
                      &  9 & $4 \cdot 10^{8}$ &  2000 & $5.0\cdot 10^{13}$ & 0 & 15&9 & 0&5 & 3&9 & 0&5 & 0&959 & 0&30 & 93\\
                      &  9 & $3\cdot 10^{11}$ &  7000 & $5.0\cdot 10^{13}$ & 0 & 14&8 & 1&4 & 4&7 & 0&5 & 0&981 & 0&28 & 65\\
                      &  9 & $4\cdot 10^{11}$ & 10000 & $5.0\cdot 10^{13}$ & 0 & 14&4 & 1&5 & 5&8 & 0&5 & 0&991 & 0&18 & 71\\
\hline
III + IV + V          &  9 & $1 \cdot 10^{9}$ &  2000 & $5.0\cdot 10^{13}$ & 0 & 16&0 & 0&6 & 2&8 & 0&5 & 0&970 & 0&41 & 75\\
\hline\hline
\multicolumn{4}{l}{$E_{\rm{stick}}$ = 0.25 eV}\\
\hline \hline
V                     &  9 & $2 \cdot 10^{12}$&  2000 & $1.3\cdot 10^{13}$ & 0 & 14&9 & 0&7 & 2&7 & 0&5 & 0&960 & 0&43 & 51\\
                      & 18 & $2 \cdot 10^{12}$&  2000 & $1.3\cdot 10^{13}$ & 0 & 10&3 & 1&0 & 2&8 & 0&5 & 0&985 & 0&46 & 51\\
                      &  9 & $2 \cdot 10^{12}$&  2000 & $5.0\cdot 10^{13}$ & 0 & 13&2 & 1&3 & 2&9 & 0&5 & 0&964 & 0&43 & 71\\
                      & 18 & $2 \cdot 10^{12}$&  2000 & $5.0\cdot 10^{13}$ & 0 & 11&1 & 1&1 & 2&6 & 0&5 & 0&978 & 0&46 & 71\\
\hline
IV + V                &  9 & $1 \cdot 10^{9}$ &  2000 & $1.3\cdot 10^{13}$ & 0 & 10&4 & 0&6 & 3&2 & 0&5 & 0&957 & 0&37 & 64\\
                      & 18 & $1 \cdot 10^{9}$ &  2000 & $1.3\cdot 10^{13}$ & 0 & 10&1 & 0&5 & 2&8 & 0&5 & 0&974 & 0&40 & 64\\
                      &  9 & $1 \cdot 10^{9}$ &  2000 & $5.0\cdot 10^{13}$ & 0 & 18&1 & 0&8 & 3&2 & 0&5 & 0&962 & 0&38 & 83\\
                      & 18 & $1 \cdot 10^{9}$ &  2000 & $5.0\cdot 10^{13}$ & 0 & 11&9 & 1&0 & 2&8 & 0&5 & 0&975 & 0&42 & 84\\

\hline \hline
Exp. \cite{Zecho:2002}     && & & & &   \multicolumn{4}{l}{17}     & \multicolumn{4}{|l}{ 4}      & \multicolumn{2}{|l}{$>$ 0.997} & \multicolumn{2}{|l|}{0.4 $\pm$ 0.2} & \\
Exp. \cite{Hornekaer:2006I} & && & & &   \multicolumn{4}{l}{}     & \multicolumn{4}{|l}{}      & \multicolumn{2}{|l}{} & \multicolumn{2}{|l|}{} & $\sim$85\\
\hline \hline

\end{tabular}
\end{table*}

\section{Sensitivity of the model on parameters \label{Sense}}
By comparing the results listed in Table \ref{Sumtable} it is clear that a combination of mechanisms IV and V
reproduces the experimental findings quite well. This section will discuss the influence of the parameter
settings on the final result. The table already showed four different parameter choices varying the diffusion
rate and the barrier for abstraction. The diffusion rate has, as mentioned before, a clear influence on the
number of formed dimers, but it  appears to have very little effect on the obtained cross sections for the
mechanisms IV and V. Lowering it to a very low value of $7.2\cdot 10^{11}$ s$^{-1}$, which turns it into a
combination of mechanisms I and IV, results in a lower cross section at low coverage. We found that reducing the
diffusion rate even further eventually leads to a completely flat cross section and a very high Pearson correlation
coefficient. Also for a pure mechanism V a higher diffusion rate gives a stronger coverage dependence of the
cross section as was suggested by Bonfanti et al. \cite{Bonfanti:2007}.

The abstraction barrier has no effect on the dimer ratio, but it influences the cross section at low coverage. It
generally gives a lower cross section for a higher barrier. For a pure mechanism I also a barrierless abstraction
was considered (see Table \ref{Sumtable}) . The cross section at low coverage did not raise to the values found
in the experiments in this case.

The parameter $B$, which controls the thermalization of the sticking atoms, has a very narrow parameter range if
only mechanism IV is considered. Lower values than $1 \cdot 10^{9}$ s$^{-1}$ result in very low sticking at low
coverage and higher values will have only a limited effect on the cross section. If mechanism IV is used in
combination with mechanism V the sticking will be increased because of the diffusion of the physisorbed atoms and
also lower values than $1 \cdot 10^{9}$ s$^{-1}$ can be used leading to a higher cross section for low deuterium
exposure and high dimer ratios. The linearity in the initial HD mass spectrometer signal is however decreased.

Adding mechanism III as a third mechanism in the simulations had a negligible effect on the results. The cross
section and initial signal curves remained unchanged within their uncertainties.

The parameter, $T_{\rm start}$, which controls the initial energy of hydrogen atoms, has a different effect on
the final results for the pure mechanism IV and the combination of IV and V. This parameter cannot be controlled
independently of $B$, since the sticking probability is constrained within a certain window (see
Fig.~\ref{therma}). For a pure mechanism IV a increased value of $T_{\rm start}$ results in a cross section which
is too low and only weakly dependent on coverage. The correlation coefficient for the initial HD signal is higher
as compared to $T_{\rm start}$ = 2000 K. This parameter combination further shows an extremely high dimer ratio.
Only combinations of mechanism were found to yield similar high values. It appear that the high initial energy is
only used to remove monomers and not for dimers to react or to desorb. The relaxation time is probably to fast
for these events to occur. If a more elaborate implementation of this mechanism was considered, where the initial
energy of atoms in dimers is high, because of their higher binding energy, the reaction from dimers would go up
and the dimer ratio go down. The results would then probably closer resemble the $T_{\rm start}$ = 2000 K
results. For a combination of mechanisms $T_{\rm start}$ has less of an effect. Here only the saturation coverage
shows a large change.

Finally we checked the influence of increasing the monomer sticking barrier to the value of 0.25 eV which is an
upper value for the sticking barrier reported by other authors \cite{Sha:2002,Sha:2002a,Morisset:2004}. This
results in a higher cross section at low coverage and a higher dimer ratio in all simulated cases. However, the
correlation coefficient of the initial mass spectrometer signal as a function of coverage goes down. In general this increase in the sticking barrier results in a better agreement with the experiments. Although the agreement of the pure mechanism V is improved much, the combination of mechanism IV and V remains the best option for both low and high values of $E_{\rm stick}$.

\section{Conclusion}

The Eley-Rideal abstraction of atomic hydrogen chemisorbed on the graphite surface has been studied via a hybrid
approach using energy barriers derived from DFT calculations as input to Monte Carlo simulations. Through
comparison with experimental data we discriminate between the contributions from different proposed Eley-Rideal
mechanisms. Good quantitative and qualitative agreement between the experimentally derived and the simulated
Eley-Rideal abstraction cross sections are found if two different Eley-Rideal abstraction mechanisms are
included. One is a direct Eley-Rideal reaction with very fast diffusion of physisorbed H atoms leading to the
formation of hydrogen dimer configurations, while the other is a dimer mediated Eley-Rideal mechanism with
increased cross section at low coverage. Such a dimer mediated Eley-Rideal mechanism has not previously been
proposed.

The effect on abstraction of fast diffusing physisorbed H atoms was first considered by Bonfanti et al.
\cite{Bonfanti:2007}, who suggested that diffusion of the physisorbed atoms could explain the high coverage
dependence of the cross section. We tested this mechanism and it was indeed found that diffusion plays an
important role, not only for the abstraction reaction but also to explain the high occurrence of dimers found by
STM measurements. However, very little is known about the exact energetic landscape in the vicinity of a
chemisorbed hydrogen atom, especially when the second atom is only weakly physisorbed. Since these weak
interactions appear to be very important, further study would be desirable.

Furthermore, we investigated the effect of steering \cite{Sha:2002,Sha:2002a} as a possibly alternative route to
reproduce the experimental findings. The results presented here show that a simple steering mechanism which just
results in increased cross sections for abstraction reactions is not sufficient to reproduce the experimental
results. This finding does, however, not rule out steering as an important mechanism. It has for example been
suggested that a coverage dependent steering effect could result from the creation of and interaction with
electron-hole pairs on the surface \cite{Hammer-Private-communications}. Further investigations of steering
should, however, consider the presence of preferred binding sites in the vicinity of adsorbed hydrogen atoms. In
the present case, sticking in a dimer position with low or no barrier is a competing reaction to the steered
Eley-Rideal reaction. Hence, the described dimer mediated Eley-Rideal mechanism offers a competing mechanism to
steering. Full quantum mechanical calculations where both channels, dimer formation and hydrogen abstraction, are
accessible, are needed to settle this case.

\section{Acknowledgments}
HC is supported by the Netherlands Organization for Scientific Research (NWO) and the Leiden Observatory. LH
acknowledges support from the Danish Natural Science Research Foundation.

%\listoffigures

\end{document}